\documentstyle[epsf]{mn}

\def\figdir{.}

\def\figname#1{\figdir/#1}

\title{Redshift of reionization: are we there yet?}
\author[Gnedin]
{Nickolay Y.\ Gnedin\\
  Center for Astrophysics and Space Astronomy, 
University of Colorado, Boulder, CO 80309, USA; gnedin@casa.colorado.edu}
\date{}
\pagerange{\pageref{firstpage}--\pageref{lastpage}}
\pubyear{}

\begin{document}

\def\dim#1{\mbox{\,#1}}

\label{firstpage}

\maketitle

\begin{abstract}
Comparison of the observed evolution of the Ly-alpha transmitted flux in the
spectra of four highest redshift quasars
discovered by SLOAN survey with the theoretical prediction for this
evolution based on the state-of-the-art numerical simulations of
cosmological reionization already allows one to constrain the redshift of
reionization to $z_{\rm REI} = 6.2 \pm 0.1{\rm s}\pm 0.2{\rm r}$, where
 systematic and  random errors are given respectively.
\end{abstract}


\section{Introduction}

In a recent paper Becker et al.\ \shortcite{bea} analysed spectra of the
four highest redshift quasars discovered by SLOAN survey. The highest
redshift ($z=6.28$) quasar showed almost no transmitted flux just blueward of
the quasar Ly-alpha emission line, and on this bases Becker et al.\ 
\shortcite{bea} concluded that reionization took place at a redshift
$z\sim6$ [a similar claim was also made by Djorgovski et al.\
\shortcite{dea} based 
on an extrapolation from a lower redshift observations].

However, one should be cautious before drawing such a conclusion. Indeed,
the observed decrease in the mean transmitted flux at $z\sim 6$ might
simply indicate a decrease in the mean ionizing intensity rather then a real
reionization of the universe. Thus, without an understanding of the
evolution of the universe around the reionization epoch, the Ly-alpha
absorption data cannot be used to constrain the epoch of reionization
[unless damping wings are observed in the absorption profiles, \cite{m98}].

Fortunately, our theoretical understanding of the process of reionization,
in part based on numerical simulations, is solid enough so that the
evolution of the ionizing intensity at these redshifts can be predicted
with a reasonable confidence level. Combining simulations with the
observational data indeed allows one to come up with a meaningful value
for the redshift of reionization.

\section{Simulations}

A set of three simulations have been performed with
the SLH code and are similar to the simulations reported in
\cite{me}. The main difference with previous simulations is that
a newly developed and highly accurate
Optically Thin Eddington Variable Tensor (OTVET) 
approximation for modeling radiative transfer \cite{ga} is used instead
of a crude Local Optical Depth approximation. The new simulations therefore
should be sufficiently accurate (subject to the 
usual limitations of numerical convergence and phenomenological description
of star formation) to be used meaningfully in comparing with the
observational data.

\begin{table}
\caption{Simulation Parameters}
\label{sim}
\begin{tabular}{@{}lccccc}
Run & 
$\Omega_{m,0}$ & 
$n$ & 
$z_{\rm REI}$ & 
$\epsilon_{\rm SF}$ &
$\epsilon_{\rm UV}/(4\pi)$ \\
A & 0.30 & 1.0  & 6.70 & 0.20 & $3\times10^{-6}$ \\
B & 0.35 & 0.95 & 5.90 & 0.15 & $6\times10^{-6}$ \\
C & 0.35 & 0.97 & 5.95 & 0.15 & $2.5\times10^{-6}$ \\
\end{tabular}
\end{table}
Parameters of the three simulations are given in Table \ref{sim}.
All three simulations included $64^3$ dark matter particles, an equal
number of baryonic cells on a quasi-Lagrangian moving mesh, and about
$100{,}000$ stellar particles that formed continuously during the simulation.
The box size was fixed at $2h^{-1}$ comoving Mpc, and the nominal
spatial resolution of the simulation was fixed at $1h^{-1}$ comoving
kpc, with the real resolution being a factor of two worse.

In all cases a flat cosmology was assumed, with $\Omega_{\Lambda,0} =
1-\Omega_{m,0}$, and {\it COBE} normalization was adopted. Notice that
because of that, a small change in the slope of the primordial power-law
spectrum $n$ makes a significant effect on the amount of the small-scale
power due to a large leverage arm from {\it COBE} scales to the tens-of-kpc
scales which are important for reionization.

Star formation is incorporated in the simulations using a
phenomenological Schmidt law, which introduces two free parameters:
the star formation efficiency $\epsilon_{\rm SF}$ [as defined by eq.\ (6)
of Gnedin \shortcite{ds}] and the ionizing radiation efficiency
$\epsilon_{\rm UV}$ (defined as the energy in ionizing photons per unit of
the rest energy of stellar particles). These two parameters are listed in
the last two columns of Table \ref{sim} for reference purposes.

The star formation efficiency $\epsilon_{\rm SF}$ 
is chosen so as to normalize the global star
formation rate in the simulation at $z=4$ to the observed value from
Steidel et al.\ \shortcite{sea}, whereas the ultraviolet radiation
efficiency $\epsilon_{\rm UV}$
is only weakly constrained by the (highly uncertain) mean
photoionization rate at $z\sim4$. The redshift of reionization
strongly depends on $\epsilon_{\rm UV}$ and is not in fact predicted in a
simulation, but can be changed over a reasonable range depending on the
assumed value of $\epsilon_{\rm UV}$. I will use this freedom to fit the
observational data from Becker et al.\ \shortcite{bea} and instead to
constrain the redshift of reionization from the observational data.

\section{A ``redshift of reionization'': what is it?}

\begin{figure}
\epsfxsize=1.0\columnwidth\epsfbox{\figname{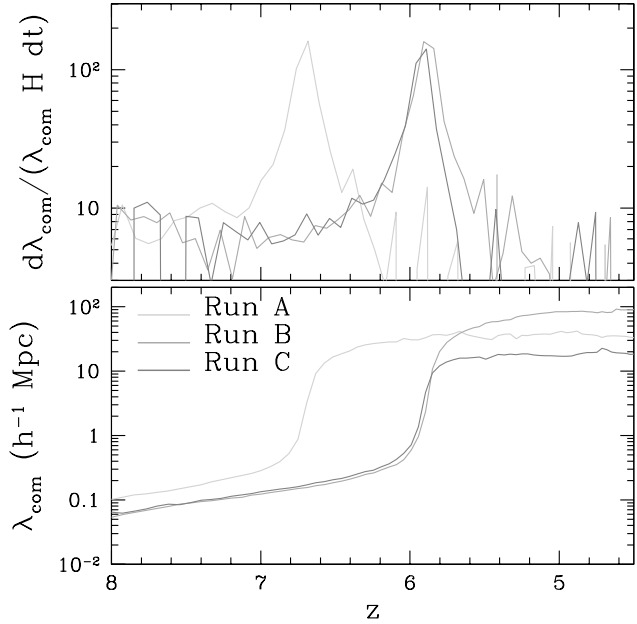}}
\caption{Mean free path to ionizing radiation (bottom) and its time derivative
(top) as a function of redshift for the three simulations from
Table \protect\ref{sim}.} 
\label{figMP}
\end{figure}
Before one can attempt to measure ``the redshift of reionization'', we
better be sure that such a quantity can be defined. The whole process of
reionization is quite extended ($\Delta z\sim5-10$), and even the
fast process of percolation of ionized bubbles
occurs over a sizable redshift interval $\Delta z\sim1$ \cite{me}.
However, one can still define the specific value of the redshift of
reionization as the moment which corresponds to the peak rate of increase
of the mean free path to ionizing radiation. As can be seen from 
Figure \ref{figMP}, the time derivative of the mean free path has a well
defined peak, which I use throughout this paper
as ``the redshift of reionization'' $z_{\rm REI}$.

\section{Results}

\begin{figure}
\epsfxsize=1.0\columnwidth\epsfbox{\figname{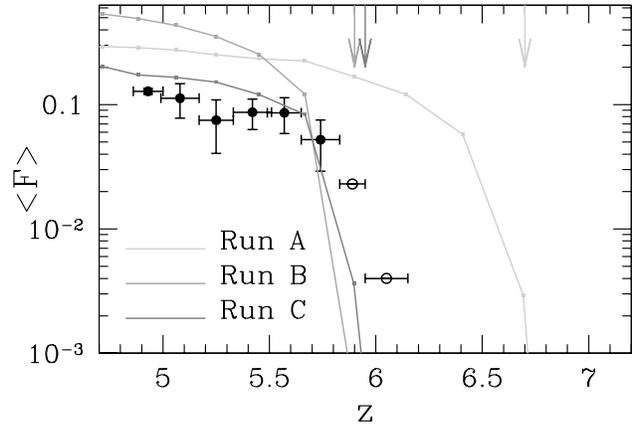}}
\caption{Mean transmitted flux as a function of redshift
for simulations A-C. Points show the observational data from
Becker et al.\ \protect\shortcite{bea}. The last two
data points (open circles) are obtained from only one quasar,
and the vertical error due to cosmic variance cannot be estimated.
Arrows mark the redshift of reionization for each simulation.}
\label{figTZ}
\end{figure}
Figure \ref{figTZ} shows the mean transmitted flux as a function of
redshift for the three simulations described above. In each case the arrow
shows the redshift of reionization, and points with error-bars are taken
from Becker et al.\ \shortcite{bea}. Each point represents an average of up
to four quasars, except for the last two points that are derived from a
single quasar and thus have no vertical error-bars. It is important to
emphasize that the measurement of the mean transmitted flux for each quasar
over an interval $\Delta z\approx 0.15$ is quite accurate, with the
intrinsic error of only $0.003-0.005$ in $\langle F\rangle$, but the
variation between different lines of sight (the so-called ``cosmic
variance'') is much greater, and it is this variation that dominates the
vertical error-bar. Thus, one (or better two) more $z=6.3$ quasars are
required in order to place the vertical error-bars on the open circles. 
However, because the sharp drop in the mean transmitted flux at $z\approx6$
is marked by two data points, it is more reliable than simply one point
from one quasar.

Simulations in Fig.\ \ref{figTZ} differ from the data points in two ways:
both the redshift evolution in a simulation and the photoionization
rate after reionization (the amplitude of the curve at low $z$) are
offset relative to the data. 
As I mentioned above, the free parameter $\epsilon_{\rm UV}$ can
be used to adjust the simulation to fit the observational data. However,
there is no guarantee that with {\it one\/} parameter I can adjust 
{\it two\/} offsets at
the same time for a given cosmological model. This fact is extremely
important because it allows one 
to actually put constraints on the cosmological
model per se, and I elaborate on this opportunity in the conclusions, but
here I am going to ignore this fact and adjust two offsets independently -
by sliding the curve both vertically and horizontally - to fit the
observational data. Because the three curves from three 
simulations have similar shapes, every simulation can thus be made to fit
the data - and, again, in reality, only a narrow range of cosmological
models will succeed in doing so. 

The reason for doing so is to obtain a constraint on
the redshift of reionization which does not depend on a (weakly
constrained) cosmological model. In addition, the simulations presented
here are rather small and numerical errors due to incomplete
convergence are substantial \cite{me}. Simulations with larger box sizes
typically have lower mean transmitted flux after reionization than small
box simulation - which implies that amplitudes of three curves are not
sufficiently accurate in Fig.\ \ref{figTZ}. The arbitrary vertical shift
of the curves can thus be considered as a ``marginalization'' over the
simulation box size.

\begin{figure}
\epsfxsize=1.0\columnwidth\epsfbox{\figname{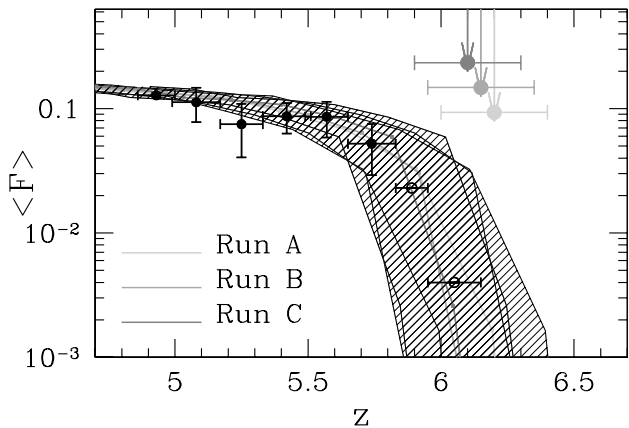}}
\caption{Mean transmitted flux as a function of redshift
for simulations A-C, fitted to the observational data from
Becker et al.\ \protect\shortcite{bea} by shifting curves
both in vertical and in horizontal direction. Shaded regions
illustrate the uncertainty due to observational errors; this uncertainty is
also given by the error-bars drawn on the arrowheads.}
\label{figTA}
\end{figure}
Figure \ref{figTA} shows the fit to the observational data for each of
the three simulations. The shaded region also gives an uncertainty of this
fit, which can be considered a random uncertainty due to observational
errors (mostly cosmic variance). Because the last two points have no
vertical error-bars, they are only partly used: they only constrain models
in the horizontal direction, and their redshift error-bars uncertainty are
somewhat arbitrarily increased by a factor of 2 (to account for the
possibility that they can be moved up and down). 
The fact that different cosmological
models have slightly different redshifts of reionization when made to fit
the observed evolution of the mean transmitted flux illustrates the
systematic uncertainty due to unknown cosmological parameters. With these
two uncertainties included, I can derive a value for the redshift of
reionization in the following form:
\begin{equation}
z_{\rm REI} = 6.2 \pm 0.1{\rm s}\pm 0.2{\rm r},
\label{res}
\end{equation}
which is the sole result of this paper.

If cosmic variance on the last two points can
be estimated, and if it is comparable to the vertical error-bar on the
$z=5.7$ point, then the random error gets reduced from 0.2 to
0.1.

\section{Conclusions}

SLOAN observations of $z\sim6$ quasars push the frontier of the observable
universe right into the epoch of reionization. Combined with the most
advanced simulations of cosmological reionization to date, the
observational data yield a rather precise measurement of the redshift of
reionization.

This measurement hinges on the assumption that the $z=6.28$ quasar (SDSSp
1030+0524) probes an average region of the universe, i.e.\ that the cosmic
variance in the measurement of the mean transmitted flux at $z=6.1$ is not
larger than a factor of 3-5. One or two more $z=6.3$ quasars are required
to confirm or refute this assumption.

A strong sensitivity of the redshift of reionization to the amount of
small-scale power offers a unique opportunity to place constraints on
the slope of the primordial power spectrum - unattainable even with the SLOAN
data on the power spectrum at megaparsec scales. For example, runs B and C
in Fig.\ \ref{figTZ} have similar star formation rates and redshifts of
reionization, but differ by a factor of 3 in the mean transmitted flux at
$z=5-5.5$ only because the slope of the primordial power spectrum $n$
differs by mere 0.02 in the two models. Sufficiently large simulations that
have numerical effects under control are currently feasible and
will eventually provide a tight constraint on the
amount of small-scale power.

\label{lastpage}

\end{document}